\documentstyle[12pt]{article}
\def\gtwid{\mathrel{\raise.3ex\hbox{$>$\kern-.75em\lower1ex\hbox{$\sim$}}}}
\def\ltwid{\mathrel{\raise.3ex\hbox{$<$\kern-.75em\lower1ex\hbox{$\sim$}}}}

\def\pc{\,{\rm pc}}
\def\kpc{\,{\rm kpc}}
\def\Mpc{\,{\rm Mpc}}

\def\cmm2{{\,\rm cm^{-2}}}
\def\cm2{{\,{\rm cm}^2}}
\def\cmm3{{\,{\rm cm}^{-3}}}
\def\gcmm3{{\,{\rm g\,cm^{-3}}}}

\def\fun#1#2{\lower3.6pt\vbox{\baselineskip0pt\lineskip.9pt
  \ialign{$\mathsurround=0pt#1\hfil##\hfil$\crcr#2\crcr\sim\crcr}}}

\begin{document}

\thispagestyle{empty}

\begin{center}


\vspace{1in}
{\Large\bf Point Sources in the Cosmic Microwave Background?\\}

\vspace{.3in}
{\large\bf Xiaochun Luo  \   \ and  \   \   Joseph Silk } \\
\vspace{0.2in}

{ Center for  Particle Astrophysics\\
301 Le Conte Hall\\
 The University of California, Berkeley, CA  94720}\\

\end{center}

\vspace{.3in}
\centerline{\bf ABSTRACT}
\vspace{0.3in}
Non-Gaussian imprints on the cosmic microwave
background radiation (CBR) sky  are within the grasp of current
experiments. A clear non-Gaussian signature would be point-like sources.
We have  examined the nature of possible point sources
 that were tentatively identified in a recent  high frequency CBR experiment
with half-degree resolution
 (Cheng et al. 1994: MSAM).
 The effects of local foreground sources, including
cold dust clouds, radio sources and the Sunyaev-Zeldovich effect due to
foreground
rich clusters are considered, and the effective spectral slopes
of these various foreground contaminations are calculated.
Radio  source emission and the Sunyaev-Zeldovich effect are  ruled out
as the explanation of the possible MSAM sources.
  Modeles are examined of extremely cold dust  clouds   which are  located
in the solar neighborhood, the interstellar medium, the galactic halo
or at cosmological distances. We estimate the cloud mass and dust grain
parameters, and in particular the grain
size distribution, that are required  in order to produce a detectable signal
in an
MSAM-type
experiment. It is shown that cold dust clouds can have an
important effect on CBR experiments only if  the clouds are nearby,  and
located
 within a disc scale height of the solar neighborhood.
Cold dust emission remains a possible source of far-infrared signal at the
level of the detected CBR fluctuations on degree scales,
 but may  be ruled out
if  the dust emissivity index satisfies $\alpha = 1.5 \pm 0.5$ .
\newpage

\section{Introduction}
Valuable information about the  early universe
and  the physical processes that generate the primordial fluctuations from
which  cosmic structure formed can be gained from measurements
of cosmic microwave background radiation (CBR) temperature
anisotropies (White, Scott \& Silk 1993).  An especially important
issue is the Gaussian nature of the temperature anisotropies. Gaussianity in
the linear regime is a generic conswequence of most inflationary theories for
the origin   of the fluctuations. However none of these models fare
particularly well at accounting for the large-scale power spectrum of density
fluctuations on all observed scales, and rather extreme solutions have been
advocated  (e.g. Peacock and Dodds  1994; Peebles 1994;
Bartlett, Blanchard, Turner and Silk 1994). Hence deviations from Gaussianity
are a possibility that can only be limited by experiment.
Although significant progress has been made in the understanding of the CBR
since
 the  detection of fluctuations
  at $7^{\circ}$ angular scale by the COBE satellite (Bennett et al. 1992;
Smoot et al. 1992; Wright et al. 1992),  the Gaussianity question
remain unresolved.   Despite the full-sky coverage achieved by COBE, the
combination of   beam smoothing and the effects of
cosmic variance  (Luo 1994; Hinshaw et al. 1994) preclude  COBE alone from
testing Gaussianity.  At the same time, while the detection of degree-scale
fluctuations intrinsic to the CBR  is a remarkable achievement,
 the datasets at these intermediate angular scales
 are still too small to carry out Gaussianity
tests,   although tests at these angular scales would certainly be  decisive
once large sky coverage is achieved (Coulson et al. 1994).

Despite all these difficulties, there are non-Gaussian imprints on the CBR sky
that are within the grasp of current on-going experiments. A clear non-Gaussian
signature would be the detection of   point-like sources. In fact, two
candidates for such sources
may have been detected by the medium scale anisotropy measurement (MSAM)
experiment.  There are serious issues of data analysis that pertain to whether
or not possible  point sources are subtracted before attempting to measure
temperature fluctuations: we do not address such issues here. Rather, we ask
the question: could possible foreground sources produce point source-like
signals
in a CBR experiment at MSAM resolution and frequency?

Various topological defects,
notably soft-domain wall bubbles (Goetz \& N\"{o}tzold 1991; Turner et al.
1991), the global
monopoles (Bennett \& Rhie 1991) or texture (Turok \& Spergel 1991) are capable
of producing
spotlike CBR anisotropies of any  desired size by choosing appropriate model
parameters.
However, before relying on topological defects as the interpretation
of candidate sources, one has to carefully filter out any foregrounds.
In carrying out the experiments, CBR anisotropy signals have to be
separated
carefully from  local foreground sub-millimeter and millimeter
radiation fields.
Three possible foreground sources are studied in this paper. These are
cold dust clouds, nonthermal extragalactic radio sources and the
Sunyaev-Zeldovich effect  in foreground galaxy
clusters. Multifrequency measurements  have previously been studied as a
ttechnique for  removing the foreground
(Brandt et al. 1994), with a focus on the point-like sources listed above.

The arrangement of this paper is as follows: an effective spectral
index for point sources is introduced in section 2, and  results for
radio sources, dust clouds and the SZ effect are presented and discussed
in section 3.
We conclude that both radio sources and the SZ effect are ruled out as a
possible explanation
of MSAM-type sources, that is to say, several $sigma$ fluctuations that are
point-like at $\sim30$ arc-min resolution and have a spectral energy
distribution that, crudely at least, is indistinguishable from that of the CBR.
Cold dust emission remains a possibility, but we find that
 this option  also may be  ruled out if a conventional value for the dust
emissivity $\alpha = 1.5 \pm 0.5$ is adopted.

\section{Effective Spectral Slope of Point Foreground Sources}
A multi-frequency technique has been used in most CBR experiments
for the purpose of removing foreground emission. The spectral
information in each pixel of the observations is used as a discriminator
between foreground and the true CBR signal. In the MSAM
experiment on which we focus in this paper, four frequency channels have been
used, $\nu = 5.6, 9.0, 16.5 {\rm \ \  and \ \ } 22.5 {\rm cm}^{-1}$.
By fitting a model for  cold dust emission to these four frequencies, we can
study the spectral features of the emission, and the range of parameters
which can still give rise to a spectrum that is close to that of the cosmic
microwave background radiation.

 Let us first define the effective spectral slope of cold dust emission.
Given the spectrum of dust emission, $S_{\nu}$, the effective
power-spectral slope $\alpha_{eff}$ is
\begin{equation}
S_{\nu} = ({\nu \over \nu_{0}})^{\alpha_{eff}} I_{B}({\nu}, T_{0}), \ \
I_{B} (\nu, T_{0}) = B_{\nu}(T_{0}) {x e^{x}\over e^{x} -1} \bigg({\delta
T\over T_{0}}\bigg)_{\rm rms},
\end{equation}
where $\nu_{0}$ is a reference frequency, which we choose it to
be $\nu_{0} = 5.6 {\rm cm}^{-1}$, $B_{\nu} (T_{0})$ is the blackbody
 radiation spectra, $T_{0} = 2.726 K$ is the CBR temperature and
  $x = h \nu/K T_{0} = 2.97 {\nu\over \nu_{0}}$. At half-degree scales,
the expected temperature anisotropy is $\bigg({\delta T\over T_{0}}\bigg)_{\rm
rms}
 \sim 1.0 \times 10^{-5}$. Thus, the expected CBR flux $I_{B} (\nu, T_{0})$ is
\begin{equation}
I_{B} (\nu, T_{0}) = 6.5 \times 10^{5} ({\nu \over \nu_{0}})^{4} e^{2.97
\nu/\nu_{0}} (e^{2.97 \nu/\nu_{0}} -1)^{-2} {\rm Jy Sr^{-1}}.
\end{equation}
By this definition, the effective spectral slope $\alpha_{eff} = 0$ for
true CBR anisotropies.

One can determine the  effective spectral slope from measuring the flux $S_{i}$
at four different frequencies $\nu_{i}$.
Let $y_{i} = log[S_{i}/I_{B} (\nu_{i},T_{0})]$ and $x_{i} =
log(\nu_{i}/\nu_{0})$,
the best fit spectral slope can be found through minimizing  $\chi^{2}$,
\begin{equation}
\chi^{2} = \sum_{i} (y_{i} - A- B x_{i})^{2}/N, \ \ N =4,
\end{equation}
here $A, B$ are two constants and the best fit slope $B$ is our effective
spectral slope $\alpha_{eff}$, which  is found to be
\begin{equation}
\alpha_{eff} = {\sum_{i} x_{i} y_{i}/N - (\sum_{i}x_{i}/N)(\sum_{i} y_{i}/N)
\over \sum_{i} x_{i}^{2}/N - (\sum_{i} x_{i}/N)^{2}}.
\label{aeff}
\end{equation}
We now discuss the flux $S_{\nu}$. The flux in each pixel is sampled
by a two-beam or three-beam chop. We will consider the situation where
in one position there is a compact source of unknown origin. By a  compact
source, we mean that the angular size of the source is smaller than the
beam width, which is $\theta_{m} = 0.425 \times 28^{\prime} = 12^{\prime}$.
 To simplify
the problem, we consider a two-beam square-wave chop (in the MSAM experiment
a sine-wave chop is used, and the position of the compact source matters if
the angular size is much less than the beam width). In this case, the observed
flux is
\begin{equation}
S_{\nu} = \pi \theta_{m}^{2} I_{B} ({\nu}, T_{0}) \delta
 + j_{\nu},
\end{equation}
where
\begin{equation}
\delta = \bigg({\Delta T \over T_{0}}\bigg) / \bigg({\Delta T \over
T_{0}}\bigg)_{rms}
\end{equation}
is the normalized temperature anisotropy. It is a random Gaussian variable
of zero mean and unit variance.
 The MSAM beam width is $\theta_{m} =
0.425 \times 28^{\prime} = 12$ arcminutes, and thus the expected CBR flux at
$\nu_{0}  = 5.6  {\rm cm}^{-1}$ is $I = 1.4 Jy$.

We parametrize the observed flux from any pixel that contains a compact source
 by a parameter $r$,
\begin{equation}
S_{\nu} = \pi \theta_{m}^{2} I_{B} (\nu, T_{0}) [ \delta + r I (\nu/\nu_{0})],
\end{equation}
where $r$ is the ratio of the contribution from the compact sources to CBR
temperature anisotropies and  $I (\nu/\nu_{0})$ is the frequency dependence
which is normalized so that $I(\nu) = 1$ at $\nu = \nu_{0}$.

As shown in Fig.1, three possible foreground sources are studied in this paper.
These are
cold dust clouds, nonthermal radio sources and the Sunyaev-Zeldovich effect  in
foreground galaxy
clusters. For each case, the flux $j_{\nu}$ from each source is discussed
in detail in the following.
\subsection{Radio Sources}
The flux is assumed to be
\begin{equation}
j(\nu) = A ({\nu\over \nu_{0}})^{B},
\end{equation}
here $\nu_{0} = 5.6 {\rm {\rm cm}^{-1}} $ is the reference frequency and
$B \approx -1$ is the spectral index of the radio emission from candidate
sources.

For the radio sources as modeled above,
\begin{equation}
r = \bigg({\theta_{d}\over \theta_{m}}\bigg)^{2} {A \over 1.5 Jy}, \ \ I
(\nu/\nu_{0}) = 0.057 ({\nu \over \nu_{0}})^{B -4}
 e^{-2.97 {\nu\over \nu_{0}} } (e^{2.97 {\nu \over \nu_{0}} -1})^{2}.
\end{equation}
Here $\theta_{d}$ is the angular size of the radio source.

\subsection{The Sunyaev-Zeldovich Effect}
The typical angular scale of a foreground rich cluster is of order
several arcminutes, which lies in the range of the MSAM sources.
The scattering of microwave photons by
hot electrons in the intracluster gas will make a rich cluster a powerful
source of submillimeter radiation. The flux density is given by (Sunyaev \&
Zeldovich 1980)
\begin{equation}
j_{\nu} = y \bigg[ x {e^{x} +1\over e^{x}-1} -4\bigg] {x e^{x}\over (e^{x} -1)}
B_{\nu} (T_{0}),
\end{equation}
where $x = h\nu/k T_{0}$ and $y = \int (kT_{e}/m_{e}c^{2}) \sigma_{T}
n_{e} dl$.

For the SZ effect, the ratio $r$ and frequency dependence  $ I({\nu\over
\nu_{0}})$ is
\begin{equation}
r = \bigg({\theta_{d} \over \theta_{m}}\bigg)^{2} {y\over (\delta T /T)_{rms}},
\ \ I ({\nu\over \nu_{0}}) = -1.41 [ \bigg({\nu\over \nu_{0}}\bigg) { e^{2.97
\nu\over
\nu_{0}} +1 \over
e^{2.97 \nu\over \nu_{0}} -1}  - 4].
\end{equation}
\subsection{Dust Emission}
  Let us consider a single dust cloud which is located a distance $D$
away from our location. The flux from the dust cloud
is (Greenberg 1968; Kwan \& Xie 1992)
\begin{equation}
j_{\nu} = {1\over 4\pi D^{2}} \int\int n_{d} (r_{d}, T)
B_{\nu}(T) \epsilon_{v} \pi 4\pi r_{d}^{2} e^{-\tau_{\nu}} dr_{d} dT,
\end{equation}
where $B_{\nu}(T) = 4.63 \times 10^{9} \nu^{3} ( \exp{4.8 \over T} \nu -1)^{-1}
{\rm Jy} Sr^{-1}$ is the brightness function of a black-body with temperature
$T$ and $\nu $ is in units of 100 GHz.    Here, $n_{d} (r_{d}, T)$ is the
distribution of dust grains  as a function of grain size $r_{d}$ and grain
temperature $T$, $\epsilon_{\nu} = \epsilon ({\lambda \over 2\pi
r_{d}})^{\alpha}$
is the emissivity of the dust grains and
  $\tau_{\nu}$ is the
optical depth of the dust emission. Since   reabsorption of
the dust emission  is expected to be negligible,
  the emission should be  optically thin so that $\tau_{\nu} =
0$.  When  the cloud consists of
dust paricles of a  single radius, and therefore of a single temperature
$T_{d}$,
  the  flux from the dust cloud is:
\begin{equation}
j_{\nu} = {N_{d}\over  D^{2}}  \pi r_{d}^{2} \epsilon_{\nu} B_{\nu} (T_{d}),
\label{s1}
\end{equation}
here, $N_{d}$ is the total number of dust particles in the cloud.

For dust emission, in addition to the parameter $r$ which denotes the ratio
of dust emission flux to the expected rms CBR flux, we have two additional
 parameters: the dust temperature $T_{d}$ and the emissivity index $\alpha$.
 $r$ and $I({\nu\over \nu_{0}})$ are given by
\begin{equation}
r = ({N_{d}/  D^{2}})  \pi r_{d}^{2} \epsilon/(\pi \theta_{m}^{2}
 (\Delta T/T_{0})_{rms}),  \ \  I({\nu\over \nu_{0}}) = \bigg({\nu\over
\nu_{0}}\bigg)^{\alpha-1}  {{ (e^{2.97
\nu/\nu_{0}} -1)}^{2} \over e^{2.97 \nu/\nu_{0}} (e^{2.97 {\nu \over \nu_{0}}
{T_{0}\over T_{d}}} -1)}.
\end{equation}

\section{Results and Discussions}
At the four frequencies $\nu_{i} = 5.6, 9.0, 16.5, 22.5 {\rm cm}^{-1}, i =1,
..., 4$ ,
which are used in the MSAM experiment, the observed flux is
parametrized as
\begin{equation}
S_{\nu_{i}} = \pi \theta_{m}^{2} I_{B} (\nu_{i}, T_{0}) [ \delta + r I
(\nu_{i}/\nu_{0})], i =1, ..., 4.
\end{equation}
The physical interpretation of $r$ is straightforward: $r/(1 +r)$ is
the percentage
of the total flux  contributed by foreground sources at $5.6 {\rm cm}^{-1}$
To determine $\alpha_{eff}$ from
Eq.(\ref{aeff}), one has to take into account the fact that
 $\delta$ is a random variable. In our calculations,  we generate 100
Monto-Carlo realizations of $\delta$  each time when computing $\alpha_{eff}$,
and
the mean and variance of $\alpha_{eff}$ for the different foregrounds are shown
 in Fig. 2 (a-f).

\subsection{The Sunyaev-Zeldovich Effect}
In Fig. 2a, the effective spectral slope
for the SZ effect is shown. Again, we find that at the 1$\sigma$ level, $r$
greater than 0.30 is ruled out based
on spectral analysis. We conclude that no more than 23\% of the
signal comes from the  SZ effect of a foreground rich cluster
 if the spectral slope lies within
$\pm 0.3$ of the CBR spectral  index.
\subsection{Radio Sources}
In Fig. 2b, the effective spectral slope
for $B = -1.0$ radio sources are shown for different ratios $r$.
Again at the 1$\sigma$ level, $r$ greater than 0.40 is ruled out based
on spectral analysis. We conclude that no more than 28\% of the
signal comes from radio emission if the spectral slope lies within
$\pm 0.3$ of the CBR spectral  index.

\subsection{Cold Dust Emission}
Estimation of the dust emission is more complicated than the calculation of the
radio source contribution and the SZ effect.
We need to specify the dust temperature $T_{d}$, the dust emissivity $\alpha$
and the geometry
of the dust grains.
The dust geometry may  be dominated by   whiskers or fractals: we will address
this point later.
In Figs. 2c, 2d, 2e, 2f, the effective spectral slopes for $T_{d} = 4 K$ cold
dust with dust emissivity $\alpha = 1.5, 1.0, 0.5, 0.0$ are shown.
The spectral slope of dust emission can be close to that of the CBR if the
dust emissivity is  small ($\alpha = 0.5, 0.0$), as shown in Figs. 2e, 2f.
Dust emission can be responsible for the putative MSAM point sources if the
dust emissivity is small.
However, if we adopt for the physical value of the dust emissivity $\alpha =
1.5 \pm 0.5$, then no more than $50\%$ of the total flux can be due to cold
dust.

 How much dust can give rise to a flux that could cause problems
for medium-scale CBR experiments?
For the time being, let us
assume that the MSAM `sources' are due to  foreground dust  emission, and we
will estimate the required dust mass.
To determine the cloud mass, we choose the  mass density of dust grains to be
$\rho_{d} = 3 g/cm^{3}$ (Hildebrand 1983). From Eqs. (\ref{s1}),
  the mass of the dust cloud is
\begin{equation}
M = {4\pi\over 3} \rho_{d} r_{d}^{3} N_{d} =
8.58\times 10^{28} ({S_{obs}\over 1 Jy}) ({\theta_{FWHM}\over \theta})^{2}
({0.1\over \epsilon})
({r_{d}\over 1\mu m})^{1-\alpha} ({D\over 1 pc})^{2}
 \nu^{-3- \alpha} [e^{{4.8\over T}\cdot {\nu}}
-1]{\rm grams}
\label{mass}
\end{equation}
Two  `sources' are reported  at $\nu = 5.6 {\rm {\rm cm}^{-1}}$
 by the MSAM experiment (Cheng
et al. 1994). The flux from one source is  $3.7 \pm 0.9$ Jy,
the other is $2.9 \pm 0.7$ Jy. Both sources are unresolved
at $\theta_{FWHM} = 28^{\prime}$.  Let us assume that the angular
size of the dust cloud is $ \theta_{d}= 14^{\prime}$ in order to minimize
the beam-smoothing effect and dust emissivity $\alpha \approx 1.5$.
 The estimated mass is
\begin{equation}
M_{1} = (1.2 \pm 0.3)\times 10^{29} ({r_{d}\over 1 \mu m})
({D\over 1 pc})^{2}{\rm grams}
\end{equation}
for the $3.7 \pm 0.9$ {\rm Jy} source and
\begin{equation}
M_{2} = (0.9 \pm 0.3)\times 10^{29}  ({r_{d}\over 1 \mu m})
({D\over 1 pc})^{2}{\rm grams}
\end{equation}
for the $2.9\pm 0.7$ {\rm Jy} source.

The dust clouds may be located in the solar neighborhood ($D < 100 \pc$),
the interstellar medium ( $100 \pc < D < 1 \kpc$), the dark halo ($ 5 \kpc < D
< 50 \kpc$) or even be at cosmological distance ($D \sim 3000 \Mpc$). Let us
look into all these possibility in detail.
\subsubsection{Dust in the  Solar Neighborhood}
Here, the solar neighborhood is loosely defined as the region centered around
the sun and within the disc scale  height ($D <100 \pc$). In the inner
solar neighbourhood, $D < 1 \pc$, the dominant heating source is the sun.
The dust mass budget is fairly small $( M \ltwid 10^{26} {\rm g})$.
 However,
the typical dust temperature is  too hot
to  account for the MSAM sources. The dust temperature is
\begin{equation}
T_{d} = [{L_{\odot}\over 4\pi D^{2}} \cdot {c^{2}\over 2\pi h}
\cdot{1\over \epsilon \zeta_{4 +\alpha}} ({c\over 2\pi a})^{\alpha}
]^{1\over 4 + \alpha},
\end{equation}
where $\epsilon \sim 0.1$ for dielectric dust grains,
$\alpha \approx 1.5$ is the gain emissivity index and $a$ is
the grain radius. For typical dust grains of radius $r_{d} = 0.5 \mu {\rm m}$,
the grain  temperature is $ T_{d} = 50 K$ for $D = 0.1 \pc$.

In the far solar neighborhood, $D \sim 100 \pc$. One expects the heat
source to be   diffuse star light.
Dust can be very cold if the  radius of the particles
is large (Rowan-Robinson 1991)  or if  they are fractals or needles (Wright
1993).
Let us first  consider fractals and needles.
 The physics of far infrared emission from fractals or needles
differs from those of the spherical dust we considered in the
previous section. Since fractals and needles have cooling times which are
shorter
than the average arrival times between incoming photons, the emission from
fractals and needles are not determined by their equilibrium temperature
but rather by their enthalpy $U(T)$,  mass spectrum
and absorption efficiency (Wright 1993).

The problem with explaining point sources with emission from
fractal dust or needles is that
 the spectra of the IR emission differs from the CBR spectrum considerably due
to
the transient heating by single photon absorption events.
The heat capacity of the dust is very small at low temperatures ($T < 10 K$),
 and absorption of a single optical
photon will increase the temperature to a very high degree unless the  fractal
or needle building block is large enough.
   The average energy of a photon in the radiation
field given by Eq. (2) is
\begin{equation}
\bar{E} = 3 k T_{s} {\zeta(4)\over \zeta(3)}  = 2.3 {\rm ev}
\end{equation}
for $T_{s} = 10^{4} K$. The enthalpy
 is
\begin{equation}
U_{g}(T) = { 4.15\times 10^{-22} T^{3.3} \over (1 + 6.51\times 10^{-3}
T + 1.5\times 10^{-6} T^{2} + 8.3 \times 10^{-7} T^{2.3})} {\rm
ergs/atom}
\end{equation}
for graphite (Guhathakutra \& Draine 1989).
The temperature increase of a building block consisting of $N$ atoms
when absorbing one photon of energy $\bar{E}$ is
\begin{equation}
\bar{T} \approx {10^{3} K \over N^{0.3}}.
\end{equation}
  Thus, $N > 10^{10}$ to avoid excessive spectral deviation from CBR,
or,  the radius of building block must be greater
 than $r_{b} \sim 0. 1 \mu {\rm m}$, which is about the size of typical
dust grains that absorb in the optical band.  For fractals of dimension D, the
emissivity is enhanced
by $(L/r_{b})^{3-D}$, where $L$ is the size of the fractal grain.
To reach an equilibrium temperature of several Kelvin, L must be
much greater than $r_{b}$. Thus, even for fractal grains the
size is large.

For clouds of large radius ($r_{d} \approx 30 \mu m$)  grains to produce
MSAM-like sources, the dust mass is
\begin{equation}
M_{d} \sim 1 M_{\odot} ({D\over 100 \pc})^{2}.
\end{equation}
Taking a typical gas-to-dust ratio $\eta = 160$, the
mass of the gas cloud is $  M = 40  M_{\odot}$
for $ D \approx 50 pc$.
The angular size of the clouds located at D = 50 pc is
$10^{\prime}$ for $l = 0.15 \pc$.
Although it is very unlikely that dust can be cooled down to a  few Kelvin
in the solar neighbourd, this window remains open for explaining the MSAM
sources.
\subsubsection{Clouds in the ISM}
In this case, the clouds are concentrated in the galactic
plane. There will be no high galactic latitude dust emission from these
dust clouds and hence there will be no effects on the  MSAM experiment, which
samples high
galactic latitudes,
even if they exist in the ISM.
\subsubsection{Dust in Cold Halo Gas Clouds}
It has been suggested that cold molecular gas clouds might be the
dominant form of galactic dark matter, distributed either in a large disk
(Pfenniger, Combes and Martinet 1994) or flattened halo
(Gerhard \& Silk 1994).  The latter model leads to the natural possibility of
cold dust clouds far from the galactic plane. The dust
temperature at the periphery of the galaxy will be as cool as
a few Kelvin even if the conventional, physically acceptable dust emissivity
is adopted.
 One initial worry is that  dust emission from such clouds
at high galactic lattitude
would have dramatic effects on the experiments that are searching for
microwave background fluctuations at millimeter and submillimeter wavelength.
We will show here that for the cloud models
given by Gerhard \& Silk (1994), there is virtually no effect on the
CBR from dust emission. In the model where the clouds are in near-hydrostatic
equilibrium, for a cloud of density $10^{4} cm^{-3}$ and temperature
100 K to satisfy the Jeans criterion requires the size and mass of the cloud to
be $L = \lambda_{J}/2 = 0.64 pc$ and $M = 85 N_{23}^{-1} M_{\odot}$.
Here $N_{23}$ is the hydrogen column density in units of $10^{23} cm^{-2}$.
Demanding the cloud-cloud  collision time scale to be at least a Hubble time
constrains the column density to be $N_{H} > 8 \times 10^{23} \cm2$.
Thus, the mass of the clouds is limited to be $M \ltwid 10 M_{\odot}$.
Even if dust dominates in  these clouds, the
flux at $5.6 {\rm cm}^{-1}$ will be $ S_{\nu} \sim 1 {\rm mJy}$. The angular
size of the dust region is $\theta_{d} \sim 10^{\prime\prime}$. After smoothing
over the beam, the observed flux will be around $1 \mu {\rm Jy}$, several
orders of magnitude smaller than  the flux from CBR temperature
fluctuations.
\subsubsection{Primeval Dust}
The last possibility of high latitude dust emission that we consider arises
from  primeval galaxies
at high redshift (Bond et al 1986).
  The angular size of the dust envelope around a primeval galaxy at redshift
$z$ is given by:
\begin{equation}
\theta = {l (1 +z) \over D_{H}}, \  \ D_{H}= 2 H_{0}^{-1} (1-{1\over
\sqrt{1+z}}).
\end{equation}
Here, $l$ is the linear size of the primeval galaxy, which is around 100
kpc. The typical angular size of the possible point sources produced
by a primeval galaxy is:
\begin{equation}
\theta = 1.14 h ({l\over 50 kpc}) ({z\over 5})  {\rm arcminute}
\end{equation}
The temperature of the CBR scales linearly with redshift $z$,
$T_{c} = 2.73 (1 +z) K$. At the periphery of the primeval galaxy with
D=50 kpc,
 the dust temperature
is $T_{d} = 24 ({z\over 5}) K$ for grain size $r_{d} = 0.5 \mu {\rm m}$,
 which is hot in the usual sense.
The observed flux density will be:
\begin{equation}
S_{obs} = 3.45 {\rm Jy} ({M/7.85\times 10^{3} M_{\odot}})
  ({2 \theta \over \theta_{FWHM}})^{2}  ({50 kpc/D_{H}})^{2} (1 +z)^{2 +\alpha}
 \end{equation}
To explain the MSAM experiment in the framework of primeval dust,
the dust mass should be:
\begin{equation}
M_{d} \sim 10^{12} M_{\odot}.
\end{equation}
This is about three orders of magnitude larger than the typical
dust mass in observed  IR galaxies. The bolometric luminosity
of the MSAM sources if they are located at cosmological distances
is $L \sim 10^{16} L_{\odot}$, much brighter than the
brightest-known IRAS galaxy F10214+4724.
It is improbable that  the MSAM sources  are primeval galaxies.

In conclusion, we have examined the nature of the putative MSAM sources
and shown that foreground radio sources or and the SZ effect in
foreground rich clusters are  ruled out as dominant contributions.
Unless the dust emissivity is unphysically small,  cold dust
emission can also be ruled out.

This work is supported in part by the NSF.

\newpage

\def\ref{\par\noindent\hangindent=2pc \hangafter=1 }
\centerline{REFERENCES}

\bigskip

\ref
Bartlett, J.G., Blanchard, A., Silk, J. \& Turner, M.S. 1994,
Fermilab-Pub-94/173-A

\ref
Bennett, C., et al., 1992, ApJ, 396, L7

\ref
Bennett, D,  \& Rhie, S., 1991, Phys. Rev. Lett. 65, 1709

\ref
Bond, J.R., Carr, B.J., \& Hogan, C.J. 1986, ApJ, 306, 428

\ref
Cheng, E.S., et al. 1994, ApJ, 422, L37

\ref
Coulson, D., Ferreira, P., Graham, P. and Turok, N. 1994, Nature, 369, 27.

\ref
Hinshaw, G., et al. 1994, COBE-Preprint-94-12

\ref
Kwan, J. \& Xie, S. 1992, ApJ, 398, 105

\ref
Gerhard, O.E. \& Silk, J. 1994, Preprint

\ref
Greenberg, J.M., 1968, in Nebulae and Interstellar Matter, ed.
B.M. Middlehurst and L. H. Aller (Chicago, The University of Chicago Press,
1968)

\ref
Luo, X.C. 1994,  Phys. Rev. D 49, 3810

\ref
Peacock, J.A. \& Dodds, S.J. 1994, MNRAS, 267, 1020

\ref
Peebles, P.J.E. 1994, Preprint, ApJL submitted

\ref
Pfenniger, D., Combes, F. \& Martinet, L. 1994 A\&A, 285, 79

\ref
Rowan-Robinson, M. 1992, MNRAS, 258, 787

\ref
Smoot, G., et al. 1992, ApJ, 396, L1

\ref
Sunyaev, R.A. \& Zeldovich, Ya, B., 1980, Ann. Rev. Astr. Astrophys. 19, 537

\ref
Turok, N, \& Spergel, D.N., 1991, Phys. Rev. Lett., 66, 3093

\ref
White, M., Scott, D. \& Silk, J. 1993, Ann. Rev. Astro. Astrophys. (in press)

\ref
Wright, E., et al. 1992, ApJ, 396, L13

\bigskip
\centerline{FIGURE CAPTION}

\bigskip
\noindent
Fig. 1: The spectral density of various sources at submillimeter range.
The solid line is the CBR anisotropies; The dotted line is SZ flux from
a rich cluster; The short dashed line is for radio sources ( spectra index -1);
The long dashed lines are for 4K cold dust emissions with different
dust emissivity $\alpha =1.5, 1.0, 0.5, 0.0$, from the top to
the bottom.

\noindent
Fig.2a: Best fit spectral slope for SZ effect.

\noindent
Fig. 2b: Best fit spectral slope for radio sources.

\noindent
Fig.2c: Best fit spectral slope for 4K dust emission with
dust emissivity $\alpha = 1.5$.

\noindent
Fig.2d: $\alpha = 1.0.$

\noindent
Fig.2e: $\alpha = 0.5$.

\noindent
Fig.2f: $\alpha = 0.0$.

\end{document}